\documentclass[aip,rsi,graphicxrsi,amsmath,amssymb,preprint,twocolumn]{revtex4-1}
\draft 

\usepackage{graphicx}
\usepackage{dcolumn}
\usepackage{bm}

\usepackage{siunitx}
\usepackage{fancyref}
\usepackage{todonotes}

\begin{document}


\title[Magnetometric Mapping of Superconducting RF Cavities]{Magnetometric Mapping of Superconducting RF Cavities}




\author{B. Schmitz}
	
	\affiliation{ 
		Helmholtz-Zentrum Berlin f{\"u}r Materialien und Energie GmbH \\ Hahn-Meitner-Platz 1 \\
		14109 Berlin, Germany
	}%
	\affiliation{Department Physik, Universit{\"a}t Siegen,\\
		Walter-Flex-Str. 3 \\
		57068 Siegen, Germany}

\author{J. K{\"o}szegi}
	\affiliation{ 
		Helmholtz-Zentrum Berlin f{\"u}r Materialien und Energie GmbH \\ Hahn-Meitner-Platz 1 \\
		14109 Berlin, Germany
	}%

\author{K. Alomari}%
	\affiliation{ 
		Helmholtz-Zentrum Berlin f{\"u}r Materialien und Energie GmbH \\ Hahn-Meitner-Platz 1 \\
		14109 Berlin, Germany
	}%

\author{O. Kugeler}
	\email{oliver.kugeler@helmholtz-berlin.de}
	\affiliation{ 
		Helmholtz-Zentrum Berlin f{\"u}r Materialien und Energie GmbH \\ Hahn-Meitner-Platz 1 \\
		14109 Berlin, Germany
	}%

\author{J. Knobloch}
	\affiliation{ 
		Helmholtz-Zentrum Berlin f{\"u}r Materialien und Energie GmbH \\ Hahn-Meitner-Platz 1 \\
		14109 Berlin, Germany
	}%
	\affiliation{Department Physik, Universit{\"a}t Siegen,\\
		Walter-Flex-Str. 3 \\
		57068 Siegen, Germany}

\date{\today}


\begin{abstract}
A scalable mapping system for superconducting RF cavities is presented. Currently, itcombines local temperature measurement with 3D magnetic field mapping along the outer surface of the resonator. This allows for the observation of dynamic effects that have an impact on the superconducting properties of a cavity, such as the normal to superconducting phase transition or a quench. The system was developed for a single cell \SI{1.3}{\giga\hertz} TESLA-type cavity, but can be easily adopted to arbitrary other cavity types. A data acquisition rate of \SI{500}{\hertz} for all channels simultaneously  (i.e., \SI{2}{ms} acquisition time for a complete map) and a magnetic field resolution of currently up to \SI{14}{\milli\ampere \per \meter \per \mu_0} = \SI{ 17}{\nano\tesla} has been implemented. While temperature mapping is a well known technique in SRF research, the integration of magnetic field mapping opens the possibility of detailed studies of trapped magnetic flux and its impact on the surface resistance. It is shown that magnetic field sensors based on the anisotropic magnetoresistance (AMR) effect can be used in the cryogenic environment with improved sensitivity compared to room temperature. Furthermore, examples of first successful combined temperature and magnetic-field maps are presented. 
\end{abstract}


\maketitle 

\section{Introduction}
Superconducting radio-frequency (SRF) cavities are enabling components of many modern particle accelerators from spallation neutron sources to CW free electrons x-ray lasers. As CW applications become more and more important, and repetition rates are increasing to obtain higher beam currents, the power dissipation in the cavity walls becomes a major limiting factor, not only from a maximum performance point of view, but also with respect to cost minimization. In the attempt to push superconducting materials to elevated performance, the degrading impact of trapped magnetic vortices must be investigated, understood, and ideally eliminated. Over the past years, several studies demonstrated in samples, in cavities and also in module-like operation how beneficial a reduction in trapped magnetic flux is. In particular, it was found that the level of trapped flux is impacted by numerous parameters, from cooldown conditions to material properties.  This laid the groundwork for further, systematic investigations \cite{Ciovati2008, Aull2012, Vogt2013a, Romanenko2014b, Vogt2015, Eichhorn2016, Posen2016, koszegi2017}.
Thus far, most of the experimental work relies on fluxgate magnetometers to measure the magnetic field during cavity operation. These sensors measure the magnetic field component in only one spatial direction and average over their length of typically \SI{20}{\milli\meter}. In addition, they are costly. Therefore, previous studies applied only a limited number of sensors, often only one. Although much insight can be gained this way, an extended, systematic approach has to include effects originating from the global distribution of the magnetic field around a cavity. Most notably, with a single 1D sensor it is impossible to distinguish between a change in absolute value and a change in direction of a measured magnetic field vector. Hence, a new affordable type of sensor had to be found which allows for high resolution magnetic field mapping at cryogenic temperatures in all three spatial directions and as a function of time. Here, we present a suitably arranged array of sensors which utilize the anisotropic magnetoresistance (AMR) effect that meets all of our requirements. \\
Furthermore, the magnetic field mapping was combined with temperature mapping\cite{doi:10.1063/1.1144532,Pekeler1996,PekelerPhD,Reschke2008} to monitor phase transitions, quench events, and local RF power dissipation during operation. Due to the flexible design of the system, additional diagnostics such as OST sensors \cite{sherlock1970,conway2008} can be added to the basic setup in future tests. The focus of this paper, however, is on the magnetic-field mapping. 

\section{System Overview} \label{sec:sysover}
\begin{figure}
\centering
\includegraphics{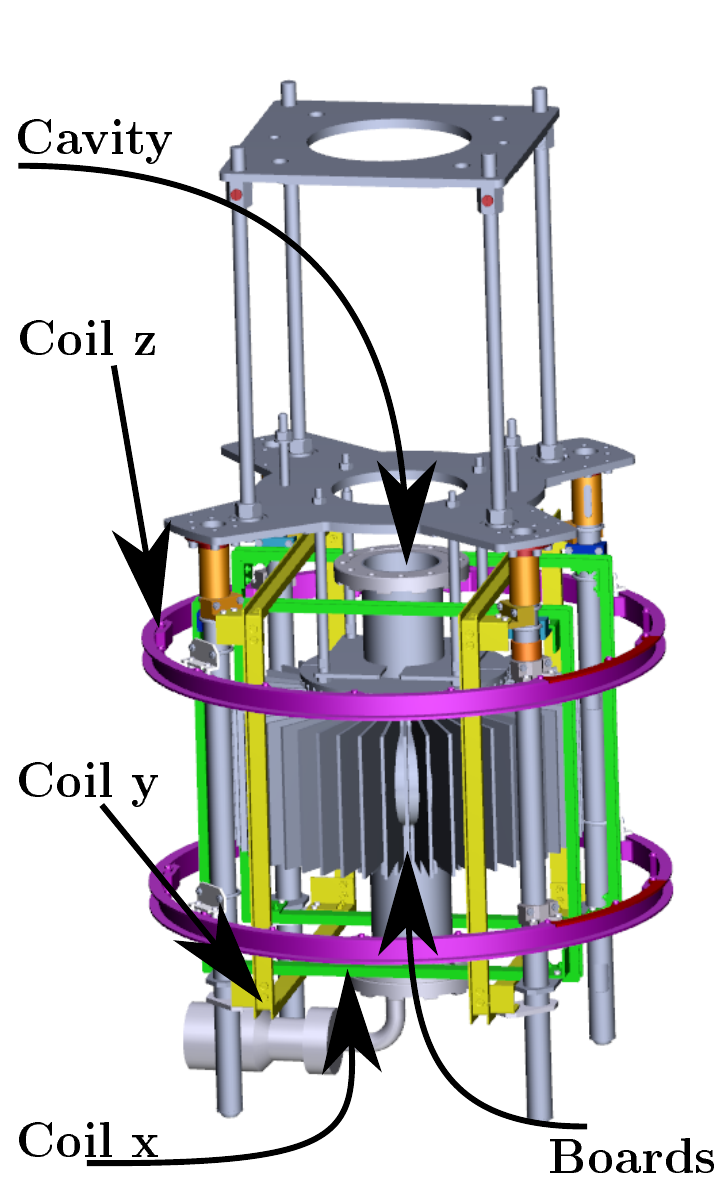}
\caption{Model of the cavity diagnostics system. It is set up as an extension to an insert of a vertical cavity test stand. A holding structure allows to attach up to 48 different temperature or magnetic field measurement boards to the cavity. Three pairs of mutually perpendicular Helmholtz coils are also attached to the insert such that the cavity is in their center. These coils allow for the generation of fairly uniform ($<$ \SI{5}{\%} deviation within the cavity volume) background fields up to \SI{200}{\micro\tesla}. Additional fixtures are integrated into the structure to include additional diagnostic devices, such as second sound sensors, not discussed in this paper.}\label{fig:basesetup}
\end{figure}

FIG.\,\ref{fig:basesetup} shows the basic setup of the diagnostics system mounted on a single-cell cavity. It is surrounded by three pairs of Helmholtz coils, one for each spatial direction. The temperature and magnetic field sensors are mounted on printed circuit boards which in turn are fixed to the beam pipes with brass holders. The holders provide 48 slots for arbitrary combinations of temperature- and magnetic-field measurement boards.

The data of all measurement channels is aquired using imc spartan devices\cite{imc}. Each device is capable of measuring 128 channels with a resolution down to \SI{10}{\micro\volt} using analog to digital converters (ADC) and amplifier cascades. Each channel has a dedicated ADC so that multiplexing is not needed. This allows for parallel sampling of all channels with a maximum sampling rate of \SI{500}{\hertz} of all boards. No additional amplifiers were used inside the cryostat. The omission of multiplexing solved the problem of having to deal with shadow signals created by capacitive crosstalk that occurs when one measures different DC voltages with a fast multiplexer\cite{knoblochPHD}. 

The following section will give a more detailed explanation of the two types of measurement boards with the main focus on the magnetometry system. Also, the underlying physics of the AMR effect and its application in the magnetic field sensor will be explained. The design of the dedicated PCBs will be shown and the handling and calibration of the sensors at cryogenic temperatures will be discussed. Finally, initial results obtained with the system will be presented.
	
\section{Magnetic Field Mapping}
\begin{figure*}[t]
\includegraphics{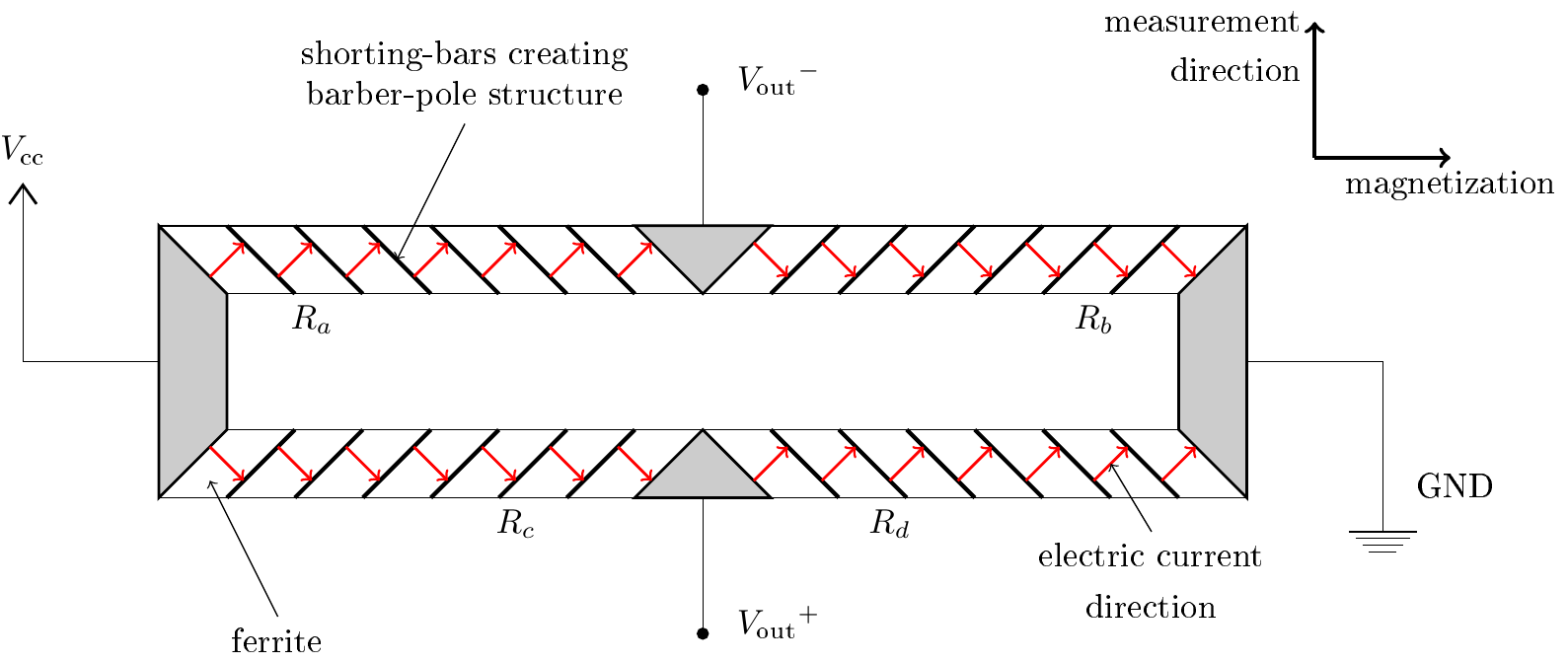}
\caption{AMR sensor consisting of four single ferrites in a Wheatstone-bridge assembly ($R_a$, $R_b$, $R_c$, and $R_d$). The sensor output is linearized by the barber pole bias which offsets the angle of current flow to \SI{45}{\degree} (indicated by the red arrows) with respect to the sensor orientation. The sensor orientation coincides with the initial magnetization in case of no applied magnetic field. The measurement direction is perpendicular to it and in-plane with the ferrite layer.}\label{fig:bridge}
\end{figure*}
Most of the recent SRF cavity experiments on flux trapping were performed with a limited number of large 1D fluxgate magnetometers. The number of used sensors as well as the spatial resolution was limited. In the attempt to upscale to a higher resolution 3D mapping system, an alternative sensor type had to be identified: first, to overcome the size limitation of fluxgates, second, to keep sensor-sensor interaction small, and third to keep the system affordable.

Based on the comparison of different sensor types \cite{CarusoMag}, magnetic sensors using the anisotropic magnetoresistance effect (AMR) were chosen. During initial testing it became evident that the used AMR sensors hold great potential: In comparison to the previously used fluxgate sensors, their area is approximately $\SI{0.7}{\milli\meter}\times\SI{0.8}{\milli\meter}$, as opposed to a typical length of \SI{20}{\milli\meter} in fluxgates which allows for a higher spatial resolution. AMR sensors exhibit a slightly better field resolution than fluxgate sensors and at a cost of a few EURO each, they are significantly less expensive. Furthermore, their sensitivity was found to improve when operated at cryogenic temperatures and they did not show mechanical damage upon several tests in cold. Their full functionality was preserved over repeated cycles between \SI{1.8}{\kelvin} and room temperature. However, these results applied only to sensors of one specific supplier. Sensors made by other manufacturers tended to fail after exposure to superfluid helium. With the sensors that were identified as cryo-compatible, 3D mapping of the magnetic field at cryogenic temperatures and during the superconducting transition becomes possible. In order to apply the mapping to superconducting cavities, PCBs with a design similiar to the existing temperature mapping boards were developed.

\subsection{The AMR Effect}\label{sec:amreffekt}
The AMR effect is a quantum mechanical effect whose macroscopic manifestation was already discovered around 1850\cite{Thomson}. It was found that the electrical resistance of a ferromagnetic material varies depending on the angle between the vector of an applied electric current and the direction of magnetization. In general, the origin for the effect lies in the combined action of magnetization and spin-orbit interaction while the magnitude of the effect is material dependent.  \cite{Campbell,FertCampbell,Ebert}. 

The AMR effect can be utilized to measure magnetic fields. Here, a ferrite with a defined preferential magnetization (easy axis) and an externally applied electric current is placed in the magnetic field that is to be measured. If the applied field is parallel to the current, the ferrite's electrical resistance takes on its maximum value. If the applied magnetic field and current are perpendicular, the value reaches its minimum. Since the change is usually in the range of a few percent, typically four AMRs are assembled in a Wheatstone bridge configuration to form a sensor as shown in FIG.\,\ref{fig:bridge}. In this configuration, the field-dependent output signal is maximized because it is no longer proportional to the absolute change in resistance $R + \Delta R$ but to the relative change
\begin{equation}
V_\text{out} = V_\text{cc}\frac{\Delta R}{R},\label{eq:wheatstone}
\end{equation}
where $V_\text{out}$ is the output voltage of the bridge and $V_\text{cc}$ is the supply voltage. 

Depending on the amplitude of the applied field, the output signal follows a $\cos^2(\Theta)$ function where $\Theta$ is the angle between magnetization and current
\begin{equation}
\frac{\Delta R}{R} \sim \cos^2\left( \Theta \right) = \left( \frac{\vec{M}'\cdot \vec{I}}{\left\vert M'\right\vert\cdot\left\vert I\right\vert} \right)^2,\label{eq:deltaR}
\end{equation}
with
\begin{equation}
			\vec{M}' = \vec{M} + \vec{H}_\text{applied}.
\end{equation}
For simplified operation, shorting bars are placed at a \SI{45}{\degree} angle on the ferrites to create a barber pole bias. Thereby, current and magnetization form a \SI{45}{\degree} angle and the dependence of the bridge output signal on the applied external magnetic field is linearized. The shorting bars are also displayed in FIG.\,\ref{fig:bridge}. Their arrangement within the Wheatstone bridge also eliminates the ambivalence of the measured field value due to the cos$^2$ dependence on the angle $\Theta$.

\subsection{Available AMR Sensors}
Reliable AMR sensors are commercially available and produced in large quantities due to their use in mobile phones and by the automotive industry. There are several manufacturers of ready-to-use sensors. For the purpose of this setup we tested three sensors of different suppliers; the KMZ51 produced by Philips, ZMY20M produced by ZETEX and AFF755 produced by Sensitec. 

As an initial test, their performance in liquid nitrogen compared to room temperature was studied\cite{ABecker2016}. The KMZ51 sensor failed operation after repeated temperature cycles and was therefore excluded. The ZMY20M and AFF755 sensors both passed the test. However, the AFF755 sensor showed a higher sensitivity. In addition, the AFF755 sensor includes coils for application of a test magnetic field (test-coil) and for magnetization recovery (flip-coil). Therefore, the AFF755\cite{DatenblattAFF} was chosen for further characterization (Sections \ref{sec:magnetization} and \ref{sec:calibration}) and later for the mapping system. During the subsequent experiments with cavities, none of the AFF755 sensors failed following operation in superfluid helium.

\subsection{Magnetometry Boards}
\begin{figure}
\includegraphics{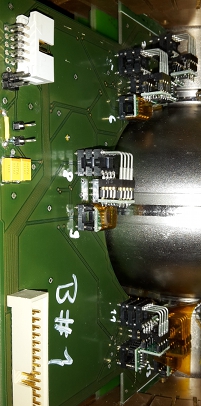}
\caption{Magnetometry board attached to a cavity. The magnetic field vector is measured at five positions using three grouped AMR sensors each.}\label{fig:MagBoard}
\end{figure}
To achieve a three-dimensional mapping, AMR sensors were positioned in groups of three facing in $r$, $\varphi$ and $z$ direction with respect to the cavity coordinate system, where $z$ is the direction of the cavity axis. Five AMR sensor groups were placed on one PCB that follows the contour of the cavity in the z-direction. One assembled board is shown as an example in FIG.\,\ref{fig:MagBoard}. 

On each board, parallel connections are used for the supply voltage of the 15 sensors. The feed line is equipped with a low pass noise filter. The test-coils as well as the flip-coils are connected in series respectively. Since the experiment is performed inside a cryostat, the supplies and the 15 signal lines are connected to the outside via one commercial 41 pin feed-through per board. Standard twisted or shielded ribbon cables were used. For future experiments with an increased number of sensors, a new multipin feed-through (1536 pins) is being built in-house. It also allows for separation of supply and signal lines.
		
In the first version of the boards, magnetic sensors were fixed with DIP plug-and-socket connections. The later version which is currently being commissioned uses directly soldered SMDs. The plug-and-socket connections are slightly more susceptible to mechanical stress but nevertheless caused no issues upon several cool down and warm up cycles ($\approx 10$). 
		
\subsection{Magnetization and Demagnetization\label{sec:magnetization}}
The AMR sensors rely on a reproducible interaction of the external magnetic field with the magnetization of the ferrites. Therefore, the magnetic domains inside each ferrite must constantly remain aligned. However, (partial) demagnetization may occur easily in sensors working at room temperature. Hence, the sensors provide integrated coils to flip the magnetization back into the original state. Thereby, the sensitivity is kept at its maximum.  

The coil can also flip the magnetization into the opposite direction. Using the sensor with both directions of magnetization allows for a correction of any sensor related voltage offset. This offset can be caused by e.g. slight variations in the resistance of the single ferrites. The output then changes according to 
		\begin{equation}
			\begin{aligned}
				V_\text{out}^+ &= &S&\times H_\text{applied} + V_\text{OS}\\
				V_\text{out}^- &= &-S&\times H_\text{applied} + V_\text{OS},
			\end{aligned}\label{eq:OS}
		\end{equation}
where the super-scripted $+$ denotes that the magnetization points into the positive direction, the super-scripted $-$ denotes the opposite direction and $S$ is the sensitivity of the sensor. Taking the difference or sum of these equations allows for the distinction of the signal from external field and offset from the sensor itself. 

\begin{figure}
\includegraphics{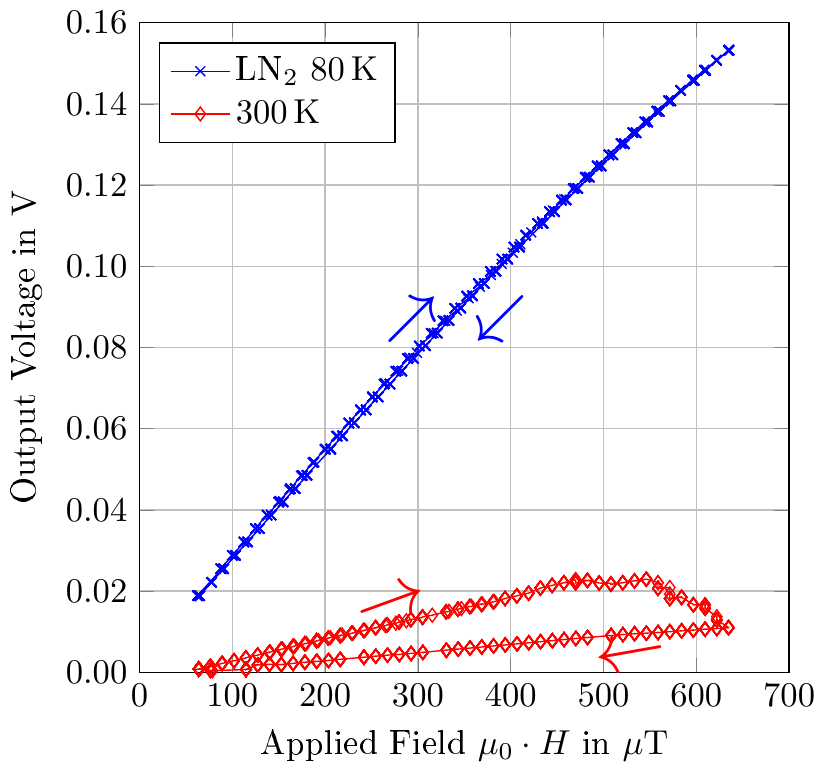}
\caption{Response of the AMR sensor at different temperatures to an applied magnetic field beeing ramped up and down once. The field was rotated \SI{150}{\degree} with respect to the measurement direction. At room temperature, demagnetization of the sensor starting at \SI{480}{\micro\tesla}. At \SI{80}{\kelvin}, no demagnetization occurred. Furthermore, the sensitivity increased at the lower temperature.}\label{fig:demag}
\end{figure}

For experiments with SRF cavities, the sensors have to be cooled down to cryogenic temperatures which impacts the material properties. FIG.\,\ref{fig:demag} shows the sensor output voltage as a function of applied magnetic field at room temperature and the same obtained in liquid nitrogen. 

The field in this experiment was rotated \SI{150}{\degree} with respect to the measurement direction. Therefore, besides the sign change, it provided a component pointing along the magnetization direction (easy axis) as well as a component in measurement direction. By comparing the two datasets, several observations can be made. First, the sensor responds linearly to the applied field at low field values, as expected. When the field is increased, the signal starts to deviate from a linear response, due to saturation of the ferrite material. At room temperature, demagnetization (migration of the magnetic domain boundaries) of the sensor sets in at $\approx\SI{480}{\micro\tesla}$. When the magnetization is partially destroyed, the signal decreases even though the applied field is still increasing because the sensitivity of the sensor decreases. Once the applied field is ramped down, the output follows the steady decrease with a constant sensitivity given by the remaining sensitivity at the highest applied field value. The original sensitivity can be recovered by using the flip-coil field to re-establish the full magnetization.

At liquid nitrogen temperature, the sensitivity of the sensor is increased given by the steeper slope which will be discussed later. Furthermore, even at high applied field no demagnetization sets in. To achieve a measurable change in the magnetization, a comparatively high field of about $\mu_0 H=\SI{800}{\micro\tesla}$ had to be applied (not shown in the figure). At lower temperatures, this value is expected to increase further. Since SRF cavity tests are performed in significantly lower field-strengths ($\mu_0 H<\mu_0\SI{80}{\ampere\per\meter}\approx\SI{100}{\micro\tesla}$), it can be concluded that no demagnetization has to be taken into account during cold tests.

On the one hand, this result simplifies performing the experiments. The use of the flip-coil can be reduced to a single magnetization set-pulse prior to the cooldown. However, on the other hand, being unable to change magnetization means that the offset of the sensor can no longer be determined easily using Equation \ref{eq:OS} which will be dicussed further in Section \ref{sec:calibration}.

\subsection{Sensor Calibration \label{sec:calibration}}
For sensor calibration, the test-coil that is integrated into the AFF755 sensor can be utilized. The advantage of this approach is that no external coil is needed and the calibration can be performed even when the boards are mounted to a cavity. The test-coil is small and close to the bridge so that neither the superconducting cavity nor the surrounding magnetic shielding impact the calibration result.

\begin{figure}
\includegraphics{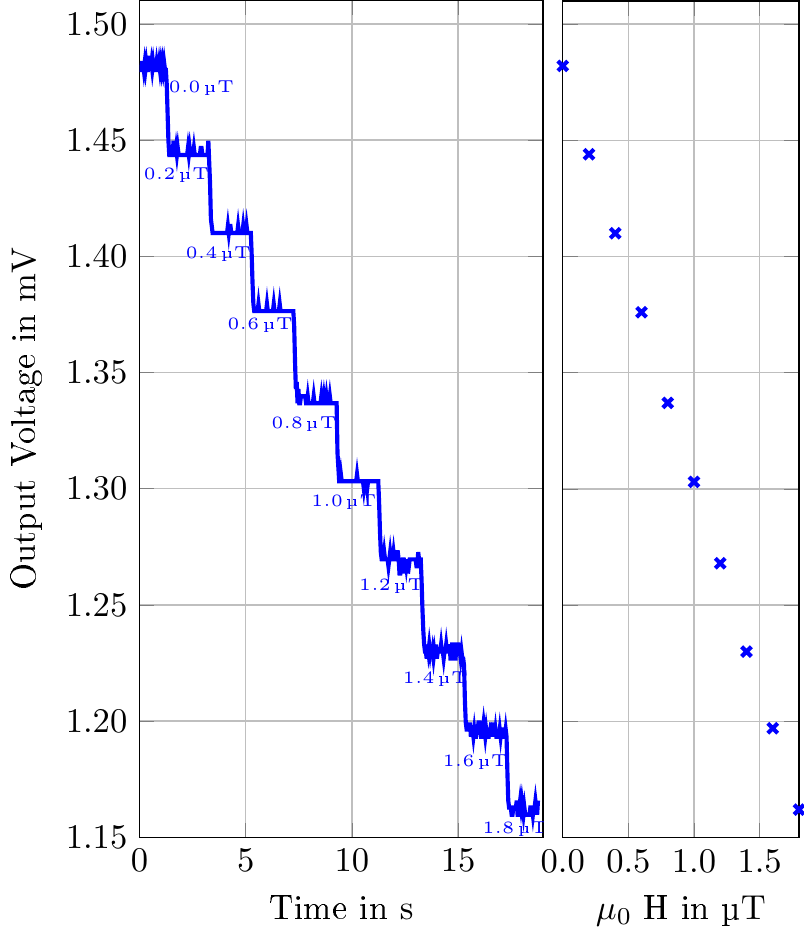}
\caption{Measurement of the sensitivity of an AFF755 sensor. The test-coil is used to apply well defined magnetic field steps to the sensor as displayed. Left: The output voltage as a response to test-coil variation. Right: Averaged output voltage vs applied field. The slope of this curve is the sensitivity.}\label{fig:sensitivitymeasurement}
\end{figure}

In addition, the magnetic field generated by the test-coil only depends on the supply current and the geometry of the coil and can be written as
		\begin{equation}
			 \mu_0 H  = C\times I,  \label{eq:hfield}
		\end{equation}
where $C$ is the geometry factor of the coil which includes, for example, its length, the number of windings, and the distance to the sensor. $C$ is assumed to be independent of temperature.

The $C$ values of each AMR sensor were determined by use of a calibrated Helmholtz coil (HC). Various HC fields amplitudes were superposed with the field of the test-coil. Each time the test-coil current was regulated to cancel the HC field which resulted in the output signal dropping to zero. A linear regression was performed of the HC field versus the test-coil current needed to cancel it. The slope of the regression yielded $C$. We found that the $C$ values of the different sensors showed only minor deviations and remained within a 5 percent range around the weighted mean: $C=$ \SI[separate-uncertainty]{0.249\pm 0.012}{\micro\tesla/\milli\ampere}.
		
Utilizing the test-coils and the determined $C$ parameters, each AMR sensor can be calibrated at any given time versus a change in flux density. To achieve this, we applied different field levels to the sensor changing them in steps, as shown in FIG.\,\ref{fig:sensitivitymeasurement}. The procedure with the HC described in the previous paragraphs also yields the offset voltage which originates from e.g.\,an asymmetry in ferrite resistances. However, a calibration during operation which is only based on the $C$ parameter cannot detect a change in the offset voltage that can occur due to inefficiencies in the magnetization procedure. Hence, at present only changes in flux density can be measured, but absolute values at a precision on the order of \SI{5}{\micro\tesla}, currently cannot be determined. Solutions for this problem of determining the offset during operation are currently under investigation. The results obtained with the complete mapping system, presented below, only show magnetic field changes relative to a starting value which is quoted as zero, but not the absolute value. 

\subsection{Temperature-dependent sensitivity\label{sec:sensitiv}}
The data in FIG.\,\ref{fig:demag} yielded an increase of sensitivity at lower temperatures. Hence, a higher output voltage can be achieved for the same supply voltage, see Equation \ref{eq:wheatstone} and the sensitivity increases. 

\begin{figure}
\includegraphics{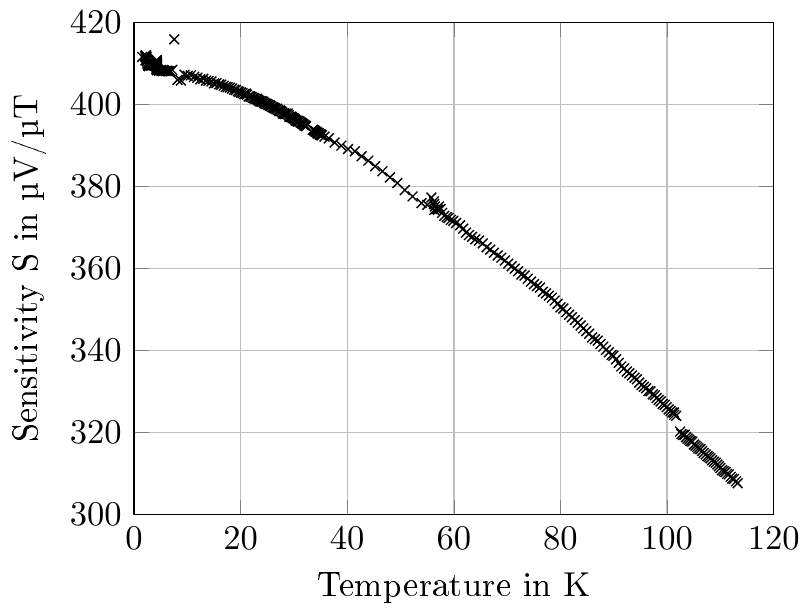}
\caption{Measured values for the sensitivity $S$ as a function of temperature $T$. It was measured from \SI{1.5}{\kelvin} up to \SI{120}{\kelvin}. Each point is the result of a stair function calibration procedure as described in FIG.\,\ref{fig:sensitivitymeasurement}.}\label{fig:sensitivity}
\end{figure}

FIG.\,\ref{fig:sensitivity} displays measured values for the sensitivity $S$ of one sensor as a function of temperature $T$. It was measured from \SI{1.5}{\kelvin} up to \SI{120}{\kelvin} by using the test-coil of the AFF755 sensor. 
Since the sensitivity of each sensor changes with temperature, it has to be calibrated individually for each operating temperature. The calibration curve has been taken for each of the 60 used sensors. Each sensitivity vs. temperature curve falls monotonically and looks similar, like the one shown in FIG.\,\ref{fig:sensitivity}. The absolute sensitivity at a given temperature, though, exhibits large differences. For \SI{1.5}{\kelvin}, mean value and standard deviation of all 60 sensors has been measured to $S=$ \SI[separate-uncertainty]{330 \pm 130}{\micro\volt / \micro\tesla}. The large standard deviation makes a careful individual calibration of each sensor inevitable for reliable operation.

\section{Temperature Mapping}
The magnetic field mapping is complemented by temperature mapping, a well established technique to measure heating at the cavity wall during operation. While a measurement of the helium losses or cavity quality factor integrates the heating of the cavity over the whole surface, temperature mapping can reveal local heating and variation of the surface resistance.

The boards used in our setup were built at DESY and were based on a system originally developed and built at Cornell University. Detailed parameters can be found in the corresponding publications \cite{Pekeler1996, PekelerPhD, Reschke2008}. 

The main change in the present system is the data acquisition which omits multiplexing as described in Section \ref{sec:sysover}. Each thermometer is read out with a maximum sampling rate of \SI{500}{\hertz}. A complete map of the cavity can therefore also be obtained every \SI{2}{\milli\second} without any crosstalk between thermometers due to multiplexing. 

\section{First Results}
\begin{figure}[htb]
\centering
\includegraphics{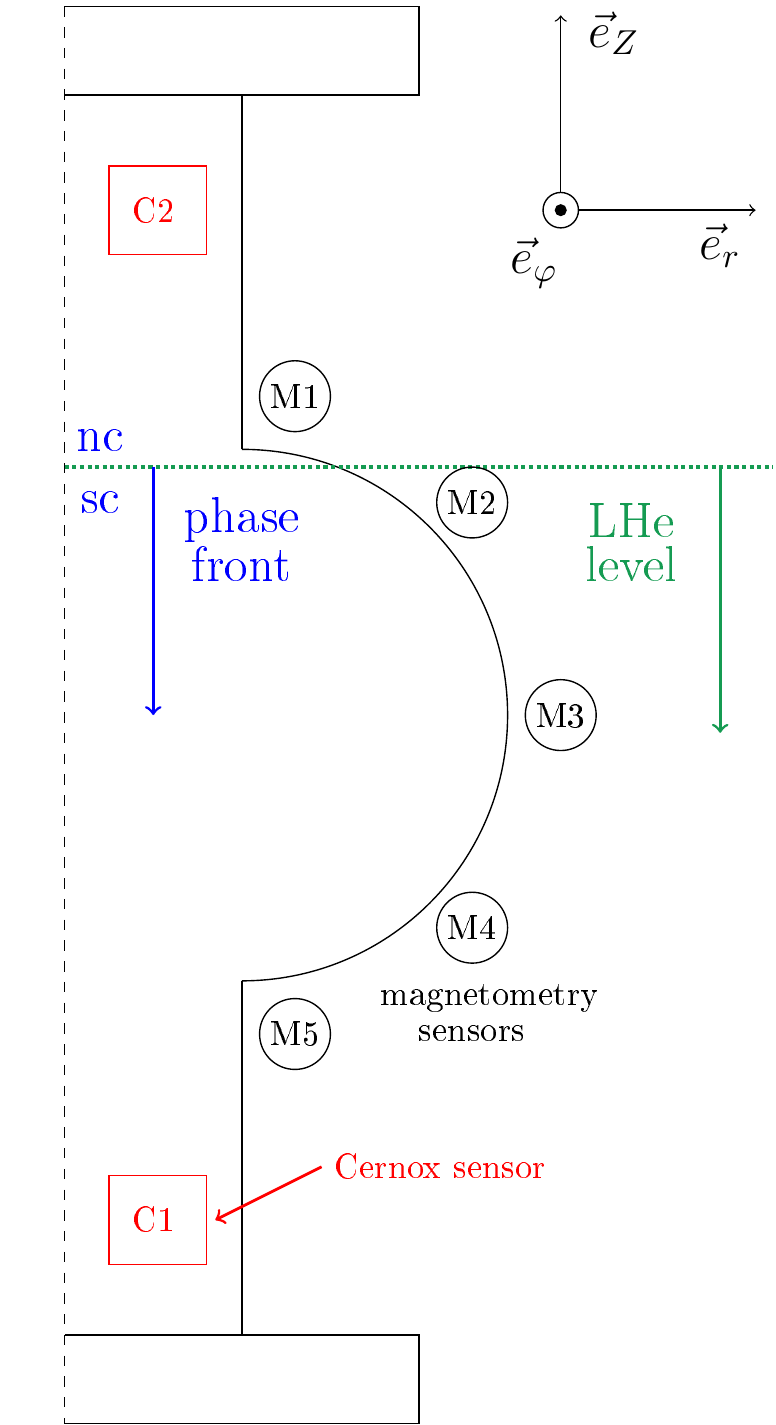}
\caption{Positions of two Cernox sensors (C) and five groups of AMR sensors (M) with a sketch of the phase front during a warm up process. Each magnetic sensor group measures the field in radial,azimuthal, and axial direction.}\label{fig:sensorCavComb}
\end{figure}
This section will give a few examples of first results obtained with the system to demonstrate the capabilities of the AMR sensors and the data acquisition system. All presented data was obtained in a vertical test assembly with a single-cell TESLA-shape cavity. FIG.\,\ref{fig:sensorCavComb} indicates the positions of the attached magnetic-field sensors.

\begin{figure}[htb!]
	\includegraphics{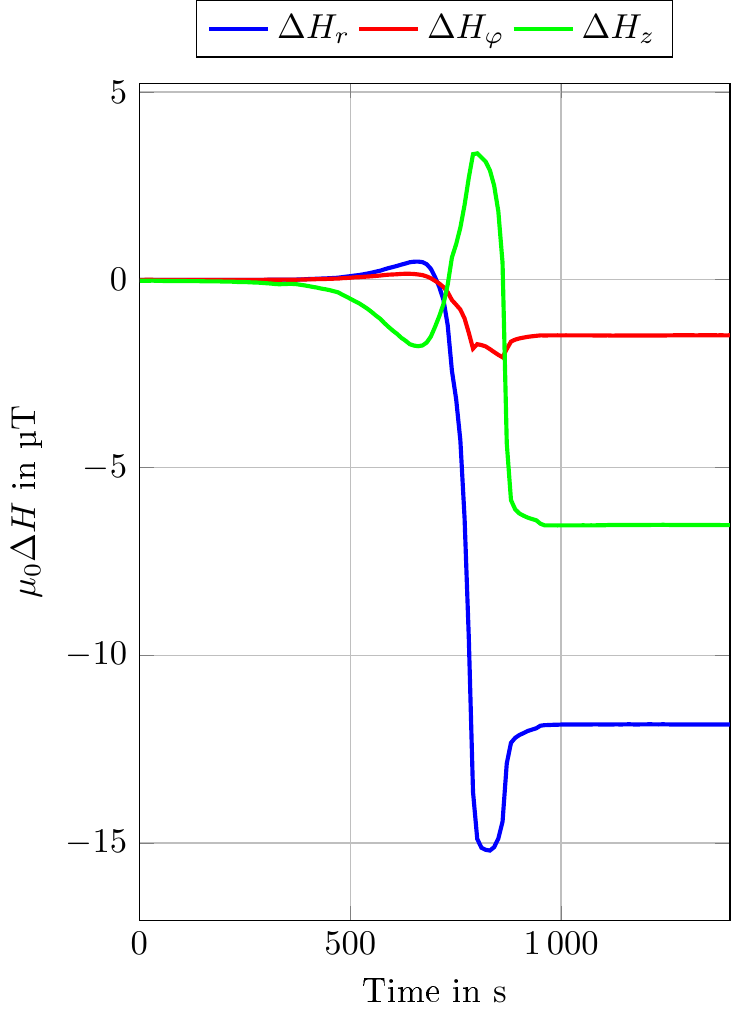}
	\caption{Release of trapped magnetic flux at the equator (M3 position).}\label{fig:warmup}
\end{figure}

\begin{figure}[htb!]
	\includegraphics{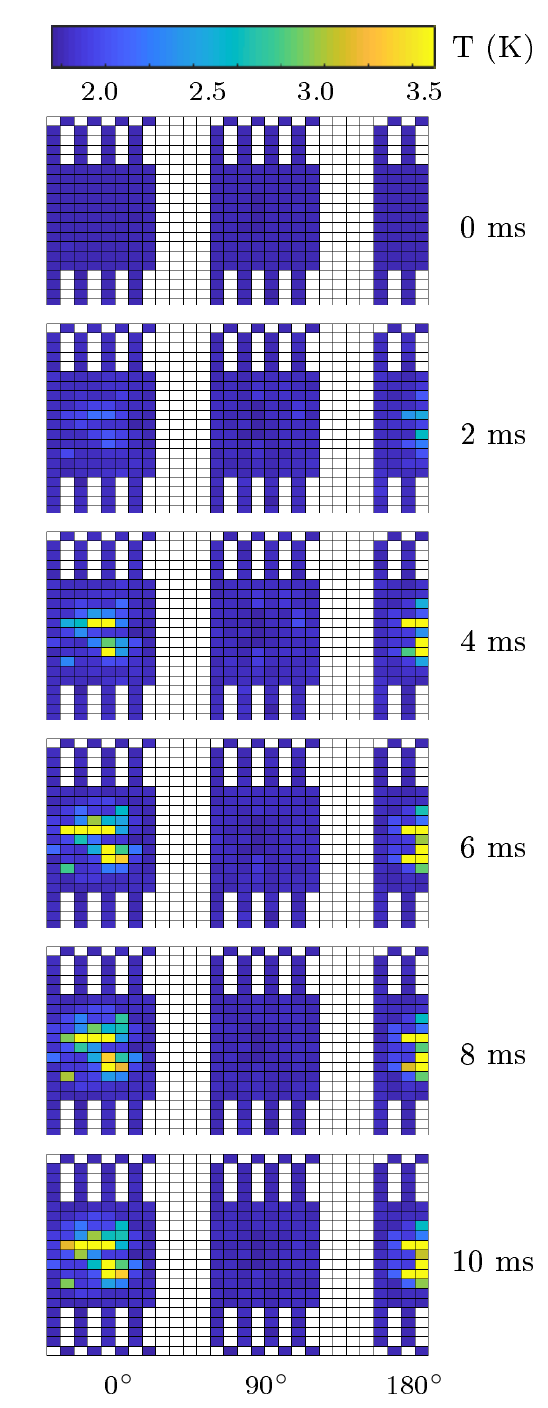}
	\caption{Temperature mapping of one half of the cavity. Two quench sponts on opposite sides of the cavity were observed. Only half the cavity was equipped with T-diagnostics, with gaps for the B-diagnostics around \SI{45}{\degree} and \SI{135}{\degree}.}\label{fig:tmapexa}
\end{figure} 
	
\subsection{Flux release during warm up}	
The first example shows the measurement of trapped-flux release during warm-up of the cavity due to the phase transition from superconducting to the normal conducting state. Since this process is slow and steady, a very smooth change in magnetic field can be observed as displayed in FIG.\,\ref{fig:warmup}. Prior to the warm up, the cavity had been cooled in $\mu_0\SI{12}{\ampere\per\meter}\approx\SI{15}{\micro\tesla}$ applied field in the direction of $\vec{e}_r$ on this specific board. Due to the not yet resolved problem of offset correction all read-outs have been zeroed at the start of the measurement. The equilibrium values towards the end of the measurement represent the magnetic flux leaving the location of the sensor and re-entering the cavity. The peaks in $\Delta H$ are a dynamic effect of the cavity's changing magnetization upon transition. Simulations \cite{Schmitz2017} indicate that this behavior can best be explained by a propagating nc/sc phase front rather than nucleation of normal conducting islands.

\subsection{Temperature Mapping without the need for Multiplexing}
A second example demonstrates the performance of the temperature mapping system. Due to the omission of multiplexing in data acquisition, the cavity can be mapped as a whole during a quench, resolving the evolution of the associated temperature map in \SI{2}{ms} steps. FIG.\,\ref{fig:tmapexa} shows a measurement where two quench spots appeared on opposite sites of the cavity surface (spaced by $\SI{180}{\degree}$ in the images) probably induced by multipacting. At the outset of the experiment, only one quench spot was present. The second spot appeared during processing that was performed on the cavity.

\subsection{Flux release during quench}		

\begin{figure}
\flushright
\includegraphics{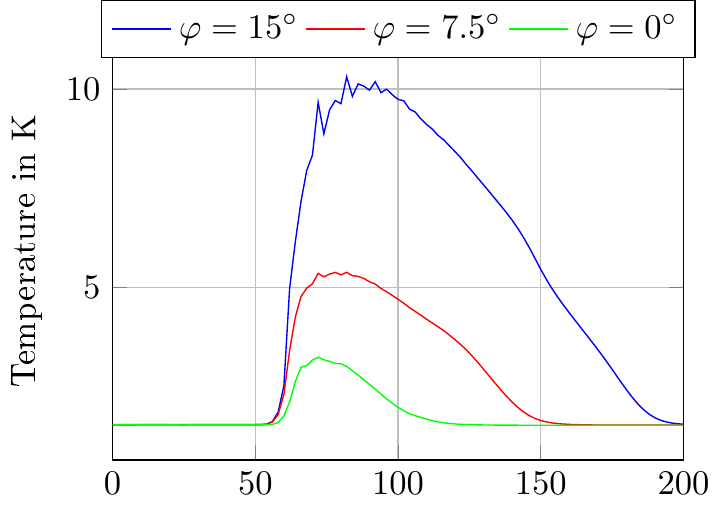} 
\includegraphics{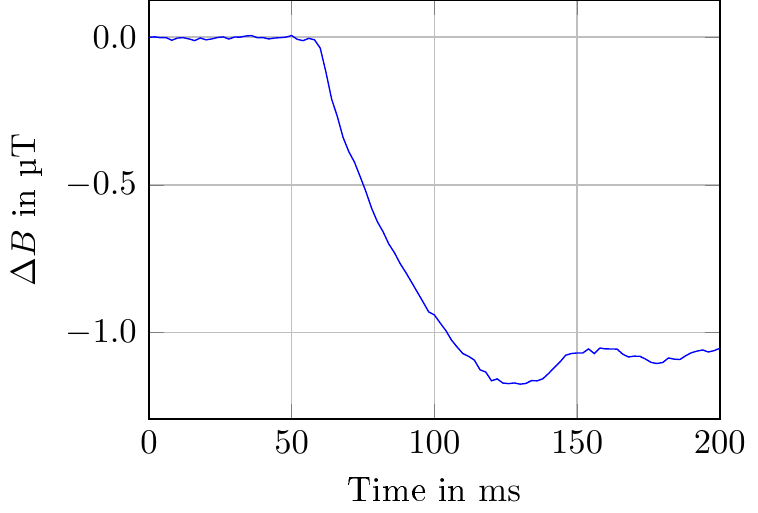} 
\caption{Combined magnetic field and temperature measurement during a quench. Top: Response of the three temperature sensors in row 13. Bottom: AMR in M2, $\vec{e}_{\varphi}$ above the quench spot. The noise at the peak of the temperature curves is a calibration artifact.}\label{fig:HTMessKomb}
\end{figure}

Finally, FIG.\,\ref{fig:HTMessKomb} shows an example for the combination of temperature and magnetic field measurement in the event of a quench. The cavity had been field-cooled in $\SI{5}{\micro\tesla}$ applied in the direction perpendicular to the board (corresponding to the azimuthal direction at the board). After the cooldown, the Helmholtz coil was switched off. 

The upper figure shows the response of three temperature-sensors close to the origin of the quench at the same $z$ position and different azimuths. The change in magnetic field measured with the magnetometry located next to the quench, as measured by the AMR in $\varphi$ direction at M2 above the equator is displayed in the bottom plot. The signal was filtered for noise at $\SI{50}{\hertz}$ and its higher harmonics.

The figure shows the correlation between measured temperature and magnetic field as the quench is expanding. The time between onset of the quench and full release of trapped flux is in the order of $\SI{100}{\milli\second}$. The dynamics of the change in magnetic field is resolved.

\section{Summary and Outlook}
We have shown that AMR sensor AFF755 can be used reliably at cryogenic temperatures. The built-in test-coil can be utilized to calibrate the sensor for relative changes in magnetic field. A resolution of \SI{17}{\nano\tesla} was achieved. The size of the averaged sensor area is approximately $\SI{0.7}{\milli\meter}\times\SI{0.8}{\milli\meter}$ which is small compared to the size of a conventional fluxgate magnetometer. Flux expulsion, trapping, and release during phase transitions of SRF cavities or during quench events can be measured. 

In combination with temperature mapping, the local dynamics of the superconducting phase front becomes accessible at a maximum data acquisition rate of \SI{500}{\hertz} which makes the presented system a powerful tool for SRF cavity diagnostics. 

Future steps will first target the absolute measurement of the magnetic field at the sub-$\mu$T level. For that purpose, the offset voltage in each sensor must be determined as a part of the calibration. One approach utilizes the curve as shown in FIG.\,\ref{fig:demag}. For higher applied magnetic field values, the sensor output deviates from the linear response. The same applies for negative test-coil currents. By measuring both sides of the curve, its symmetry point can be determined. The distances of this point from the zero on the ordinate and on the abscissa hold the information on the contributions to the offset due to internal sensor asymmetries or external field. A calibration sequence implementing this approach is currently being investigated.

Furthermore, a possible degradation due to radiation must be investigated. AMR sensors have been tested for aerospace application and achieved a radiation hardness of over \SI{10}{\kilo\gray}\cite{radiation}. However, this value was obtained in experiments that also included the readout electronics exposed to the radiation -- which is not the case for our setup. Since the AMR sensors themselves are passive devices a much higher radiation limit is to be expected. During the tests presented we found no limitation so far. However, the radiation hardness will be included in further testing. In addition, the degradation of all materials during frequent testing will be studied and the mechanical components as well as the circuitry will be optimized. 

Additionally, it is planned to integrate temperature and magnetic sensors on a single board in order to eliminate ``blind spots'' in the azimuthal direction. Finally, other types of magnetic field sensors are investigated for SRF application in the medium term. The use of GMR- utilizing the giant magnetoresistance effect or even TMR sensors utilizing the tunnel magnetoresistance effect will be evaluated because they feature different sizes as well as different sensitivities.

\bibliography{magnetometry}

\end{document}